# Enhancing TiFe Alloy Activation for Hydrogen Storage Through Al, Cr, Co, and Cu Substitutions as a Step Towards Future Recycling

Francesca Garelli,[a] Erika Michela Dematteis,[a]* Vitalie Stavila,[b] Giuseppe Di Florio,[c] Claudio Carbone,[c] Alessandro Agostini,[c] Mauro Palumbo,[a] Marcello Baricco,[a] Paola Rizzi[a]

[a] Department of Chemistry and NIS, INSTM, University of Turin, Via Pietro Giuria 7, 10125 Torino, Italy

[b] Sandia National Laboratories, 7011 East Ave, Livermore, CA 94550, United States

[c] ENEA: Italian National Agency for New Technologies, Energy and the Environment, Italy

*Corresponding author

Erika Michela Dematteis

E-mail address: erikamichela.dematteis@unito.it

## Graphical Abstract

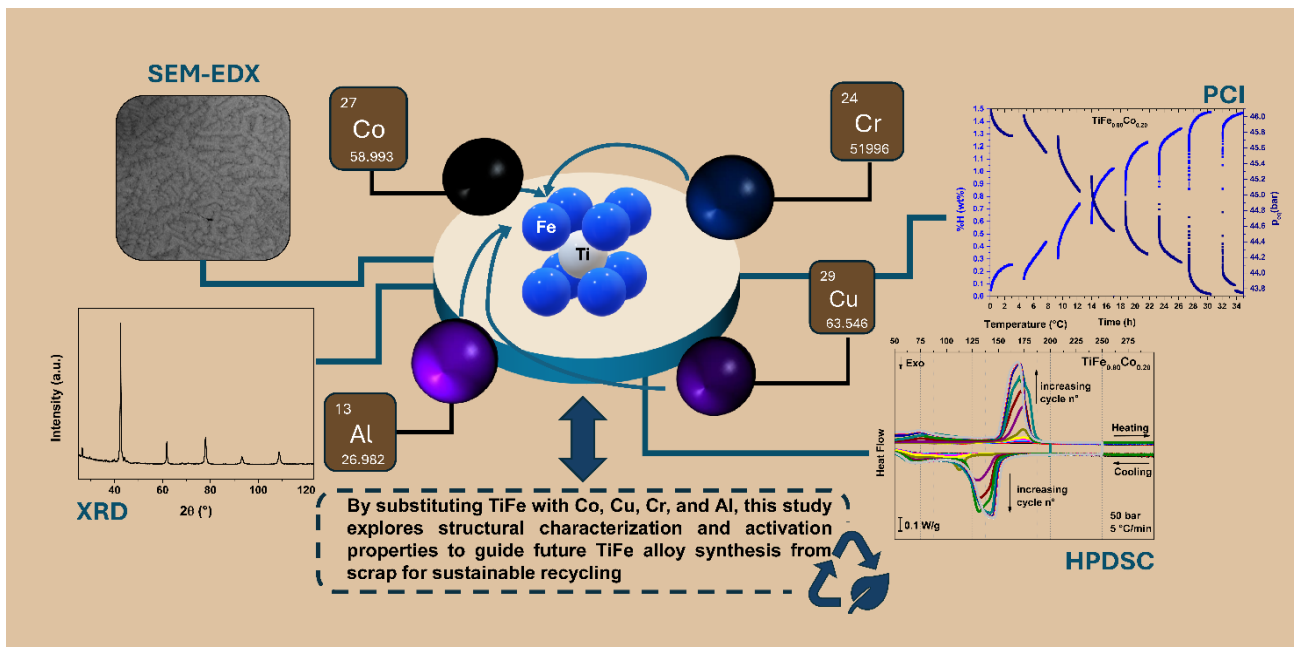

**Abstract**

This study investigates the activation behavior of TiFe$_{0.80}$-X$_{0.20}$ (X = Co, Cu, Cr, Al) alloys to identify the most effective materials for producing hydrogen storage alloys from recycled sources in view of a circular economy perspective. Activation was tested using two methods: a Sievert Volumetric Apparatus at room temperature and 64 bar of hydrogen, and high-pressure differential scanning calorimetry with 50 bar hydrogen under thermal cycles up to 400 °C. Activation properties were analyzed by assessing time for incubation and for full charging, that are influenced, respectively, by surface and bulk diffusion of hydrogen. Results showed that Cr-substituted alloys are rapidly activated, due to the presence of TiCr$_2$ compound, while Al-containing alloys absorbed hydrogen immediately. In contrast, Co- and Cu-substituted alloys required extended activation times, due to less quantity of secondary phases and limited diffusion channels.

# Introduction

According to the International Energy Association, electricity consumption per capita at the global level increased by 58% during 1990-2018 [1]. On the other hand, thanks to the increase in use of renewable energies, the $CO_2$ emission per capita raised much less at the global level (i.e., 14%). These values are very different considering different areas, as demonstrated by the difference in energy consumption and $CO_2$ emission among Europe (5.60 MWh/capita*year and 5.72 $tCO_2$/capita*year), Africa (0.57 MWh/capita*year and 0.98 $tCO_2$/capita*year) and Asia (6.04 MWh/capita*year and 8.58 $tCO_2$/capita*year) [1]. The difference in the $CO_2$ emission in different areas is the result of the diverse penetration of renewable sources in energy production [2].

The use of hydrogen as an energy vector has garnered global attention as a key element for addressing the geopolitical, social and environmental challenges stemming from the substantial increase in energy consumption. Hydrogen technologies represent a potential solution for storing surplus electricity generated by intermittent renewable sources, which exhibit significant temporal and geographical fluctuations, in dispatchable fuels. This alternative offers numerous advantages, including capacity for prolonged storage, and a superior energy-to-weight ratio in comparison to alternative fuels. However, there exists a converse relationship in terms of energy-to-volume ratio, representing a drawback for hydrogen storage. This limitation can be addressed by enhancing energy density by constraining hydrogen in a reduced volume [3]. Various methodologies can be employed for hydrogen storage, encompassing compressed gas, liquid and solid-state. Amid the array of plausible hydrogen storage methods, solid hydrogen carriers emerge as a promising solution due to their ability to achieve substantial volumetric densities (>80 kg/m$^3$) relative to compressed (≈30 kg/m$^3$) or liquid hydrogen (≈40 kg/m$^3$) [4]. Furthermore, the handling of hydrogen in metal hydrides proved to be a safer alternative compared to liquid or compressed gas [5].

The most widely recognized compounds for hydrogen storage applications are TiFe and its modifications, achieved by substituting Ti or Fe with other metals [2,6]. Despite its promising hydrogen storage capacity, challenges related to material activation and sloped hydrogen absorption/desorption plateau pressures hinder the use of pure TiFe [7–12]. In recent years, various attempts to enhance the hydrogen storage properties of TiFe have primarily focused on elemental substitution. The substitution of elements forming high stability hydrides (e.g., Nb, Zr, V) to the Ti site can modify the strength of the hydrogen-metal bonds, while the incorporation of elements forming low stability hydrides (e.g., Cu, Ni, Co, Mn, Al) into the Fe site can improve the activation process [6].

TiFe alloys (atomic radii, $r_{Ti}$ = 0.147 nm and $r_{Fe}$ = 0.126 nm) with different additions have been widely investigated [13]. The substitution of Fe with Co ($r_{Co}$ = 0.125 nm) decreases the first plateau pressure and reduces hydrogen absorption capacity, forming only the monohydride [13–16], while in the as-cast state, it increases the second plateau pressure [13,17]. Cu substitution ($r_{Cu}$ = 0.128 nm) expands the cell parameter and lowers the first plateau pressure. This substitution and the formation of the $Fe_2O_3$ phase improve the activation [13,18,19]. Cr ($r_{Cr}$ = 0.128 nm) promotes $TiCr_2$ formation, accelerating activation, stabilizing the first plateau, but shortening the second one [13,20,21]. TiFe-Cr alloys are hard and brittle, facilitating pulverization and improving kinetics, though its effect on hysteresis remains debated [22,23]. Al substitution ($r_{Al}$ = 0.14317 nm) enhances kinetics, but leads to sloped plateaux, increases equilibrium pressures, reduces absorption capacity, and inhibits γ-hydride formation, while lowering hysteresis due to valence electron differences [13,16,24,25].

This paper explores the activation process of TiFe alloys with the addition of a third element, analyzing compositions such as $TiFe_{0.80}Co_{0.20}$, $TiFe_{0.80}Cu_{0.20}$, $TiFe_{0.80}Cr_{0.20}$, and $TiFe_{0.80}Al_{0.20}$. The study investigates how these elements influence the alloy's crystal structure, microstructure, and key parameters, like hydrogen diffusion coefficient and activation energy for the diffusion. Notably, the Cr-containing alloy exhibits superior absorption and activation properties, enhanced by a secondary phase that facilitates hydrogen uptake. Additionally, annealing plays a crucial role in tuning the alloy's properties, as revealed in both Al- and Cr-modified samples.

As previously mentioned, TiFe alloys face challenges related to material activation, so using scrap in their production will introduce additional elements beyond Ti and Fe that could impact activation, making it essential to understand these effects. Therefore, the activation process is further analyzed by considering the numerous factors affecting it, from structural changes to external conditions. Ultimately, these insights pave the way for a clearer understanding of how to efficiently synthesize valuable alloys from scraps, where the presence of additional elements can significantly impact the final properties.

## Materials and methods

### Synthesis and processing

An amount of 10 g of each alloy was prepared from the parent elements by arc melting (Edmund Bühler Arc Melter D-7400) under vacuum ($10^{-4}$ bar). Raw materials, provided by Alfa Aesar and Thermo Scientific, are Ti Grade 1 (minimal purity 99.99%), Fe (minimal purity 99.97%), Al (minimal purity 99.999%), Co (minimal purity 99%), Cr (minimal purity 99.99%), Cu (minimal purity 99%). Samples were obtained by weighing the three elements in the correct proportions, based on the desired chemical composition: $TiFe_{0.80}Co_{0.20}$, $TiFe_{0.80}Cu_{0.20}$, $TiFe_{0.80}Cr_{0.20}$ and $TiFe_{0.80}Al_{0.20}$. In the case of Cr and Al additions to TiFe, samples were annealed in a furnace (Carbolite STF 16/180 tube furnace) for 5 days at 1000 °C under dynamic vacuum ($10^{-3}$ bar), leading to annealed $TiFe_{0.80}Cr_{0.20}$ and annealed $TiFe_{0.80}Al_{0.20}$.

The resulting ingots were crushed in air with a hammer to chunks, and part of them were finely grounded with a mortar and pestle to obtain fine powder in air.

## Characterization

### Powder X-ray diffraction

In order to do a structural characterization of the as-synthesized alloys, the synthesized powders obtained from the chunks were analyzed with X-ray Diffraction (XRD) analysis and subsequently stored in air at ambient conditions. The XRD analysis was conducted using a Malvern Panalytical Empyrean Diffractometer in the Bragg-Brentano configuration, equipped with Cu-K$\alpha$ radiation ($\lambda$=1.5406 Å) and a 1Der detector that reduces the fluorescence interference.

XRD patterns were collected in air at room temperature, with an acquisition time of 240 seconds per step, spanning from 5° to 124° in 2$\theta$. Qualitative analyses were carried out using the X-Pert High Score software, and a Rietveld refinement was performed for quantitative and crystal structure analysis using the Maud software [5]. $LaB_6$ standard was used to obtain a standard XRD pattern for the Maud software calibration.

### Scanning Electron Microscopy

The samples chunks were hot embedded in a conductive resin and polished for Scanning Electron Microscopy (SEM) analysis to investigate the microstructure of the as-cast and annealed samples. The examinations were carried out using a Field Emission Gun - Scanning Electron Microscopy (FEG-SEM) instrument, specifically the Tescan 9000. Energy Dispersive X-ray Spectroscopy (EDX) elemental measurements were performed on the embedded samples at 20 keV and 1 nA.

### Hydrogen sorption properties

Activation properties were assessed by Pressure-Composition-Isotherm (PCI) and High-Pressure Differential Scanning Calorimetry (HPDSC) measurements, using a PCT-Pro by Setaram and a DSC 204 HP Phoenix by Netzsch, respectively.

For the PCI measurements, approximately 2 g of sample were loaded into a stainless-steel sample holder in air, then evacuated under primary vacuum at 300 °C for 1 h and activated by exposing it to pure gaseous hydrogen (Nippon gases, 99.9999 %) at room temperature. Each absorption cycle was done at 64 bar of hydrogen (45 bar in the case of Co alloy) and room temperature, followed by a 30 min desorption cycle at room temperature and a 30 min evacuation using a rotary pump ($10^{-2}$ bar) also at room temperature.

HPDSC measurements were carried out by loading, inside the glove box, the samples (approx. 60 mg) into Al crucibles, which contained a hole in the top lid to allow $H_2$ to enter in the sample holder. Once the crucibles were loaded into the instrument chamber, the Ar atmosphere was removed using a primary vacuum at room temperature.

The HPDSC activation process involved a heat treatment under dynamic primary vacuum, where the temperature gradually increased at a uniform rate of 5 °C/min, starting from room temperature and reaching 400 °C, followed by 30 minutes isotherm at 400 °C and a subsequent cooling down to room temperature was carried out at 5 °C/min.

Following the heat treatment, the sample was held at 40 °C for 3 h, while being exposed to 50 bar of hydrogen for activation. Afterwards, the sample underwent several heating-cooling cycles at a rate of 5 °C/min from 40 °C to 400 °C under a static hydrogen pressure of 50 bar. During heating, endothermic processes were typically observed, resulting in desorption peaks, while during cooling, exothermic processes were generally seen, leading to absorption peaks. This cycling process was repeated multiple times, and the sample was considered fully activated when the absorption and desorption behavior became reproducible for at least five consecutive cycles.

## Results and discussion

### Chemical and structural characterization

**Figure 1** shows the backscattered electron (BSE) images of all the samples, providing a visual overview of the matrix and secondary phases present in the TiFe-X alloys. In these images, the matrix phase is the dominant feature, with secondary phases appearing as distinct regions. The quantity of secondary phases is minimal compared to the matrix in most samples. These phases, which depend on the third element (X) added to the alloy, can be seen in the BSE images as areas with different grey tones, corresponding to different compositions. In Al-based as-cast alloy, the matrix itself exhibits two distinct grey tones, indicative of compositional inhomogeneity present after the synthesis.

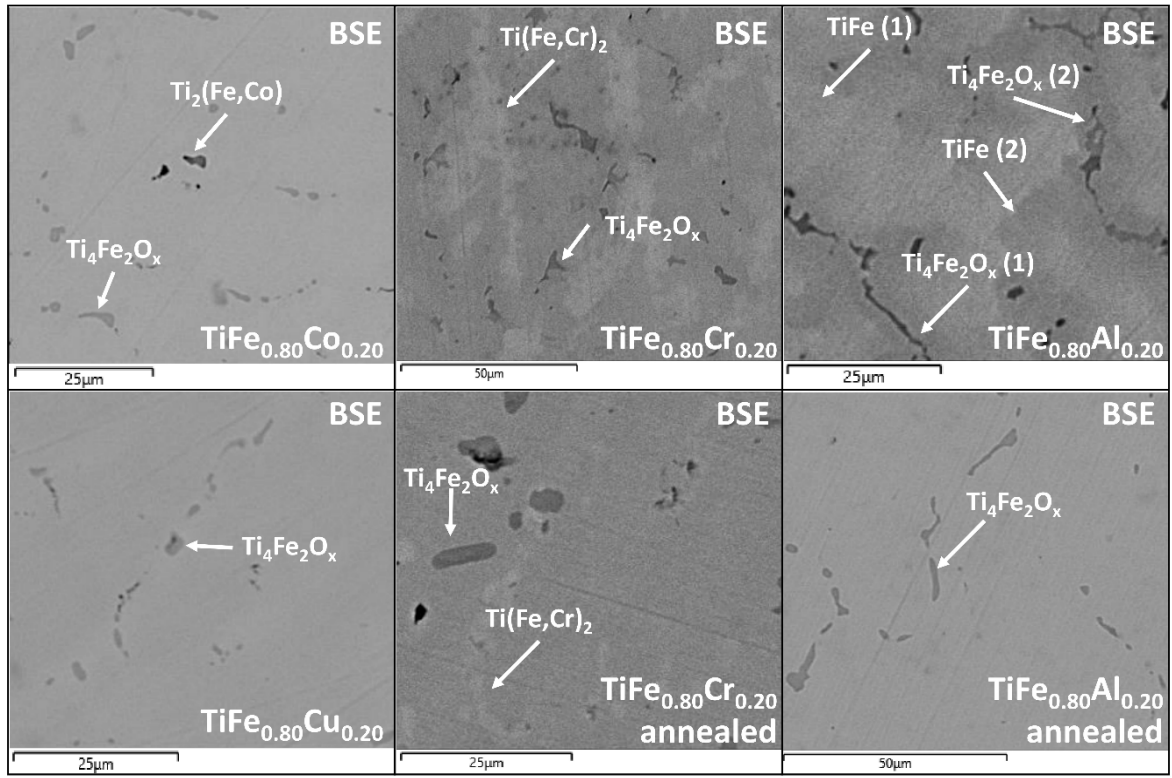

**Figure 1: BSE metallographic images of all samples by SEM.**

**Table 1** presents the results of the EDX analysis, showing the compositions of the different phases present in the investigated alloys. The standard deviation was calculated using its formula, based on the square root of the variance of the measured values.

| Sample | Average Composition | | | Phase | Phase Composition | | |
|---|---|---|---|---|---|---|---|
| | Ti (at%) | Fe (at%) | X (at%) | | Ti (at%) | Fe (at%) | X (at%) |
| TiFe$_{0.80}$Co$_{0.20}$ | 51.3±0.2 | 39.0±0.1 | 9.7±0.2 | TiFe | 51.3±0.3 | 38.5±0.9 | 10.2±1.1 |
| | | | | Ti$_4$Fe$_2$O$_x$ | 60.7±1.7 | 33.7±1.2 | 5.7±0.5 |
| | | | | Ti$_2$(Fe,Cr) | 72.3±5.3 | 23.5±4.6 | 4.3±0.9 |
| TiFe$_{0.80}$Cu$_{0.20}$ | 51.1±0.1 | 39.5±0.6 | 9.4±0.5 | TiFe | 50.8±0.2 | 40.6±0.3 | 8.7±0.4 |
| | | | | Ti$_4$Fe$_2$O$_x$ | 63.0±1.8 | 29.4±1.4 | 7.6±0.8 |
| TiFe$_{0.80}$Cr$_{0.20}$ | 51.2±0.2 | 38.4±0.5 | 10.4±0.7 | TiFe | 52.4±0.4 | 40.7±0.6 | 6.9±0.9 |
| | | | | Ti$_4$Fe$_2$O$_x$ | 69.0±2.0 | 21.4±1.9 | 9.6±0.3 |
| | | | | Ti(Fe,Cr)$_2$ | 44.2±1.1 | 39.2±1.8 | 16.6±0.9 |
| TiFe$_{0.80}$Cr$_{0.20}$ Annealed | 51.0±0.2 | 39.1±0.2 | 9.9±0.1 | TiFe | 51.6±0.1 | 40.2±0.2 | 8.2±0.3 |
| | | | | Ti$_4$Fe$_2$O$_x$ | 61.3±1.8 | 29.7±2.4 | 9.0±0.6 |
| | | | | Ti(Fe,Cr)$_2$ | 41.6±0.6 | 31.7±0.8 | 26.7±1.4 |
| TiFe$_{0.80}$Al$_{0.20}$ | 50.6±0.1 | 37.8±0.1 | 11.6±0.1 | TiFe (1) | 48.7±1.0 | 41.1±2.3 | 10.2±1.3 |
| | | | | TiFe (2) | 50.3±1.1 | 37.2±3.0 | 12.4±2.2 |
| | | | | Ti$_4$Fe$_2$O$_x$ (1) | 61.8±3.1 | 25.8±2.6 | 12.4±0.5 |
| | | | | Ti$_4$Fe$_2$O$_x$ (2) | 69.3±1.5 | 16.7±1.3 | 14.0±0.2 |
| TiFe$_{0.80}$Al$_{0.20}$ Annealed | 50.5±0.2 | 37.8±0.1 | 11.8±0.1 | TiFe | 50.5±0.1 | 39.5±0.2 | 10.1±0.2 |
| | | | | Ti$_4$Fe$_2$O$_x$ | 65.6±0.3 | 22.8±0.3 | 11.7±0.3 |

**Table 1: Sample list, nominal composition, observed phases and elemental analysis of phases in at%, with the respective standard deviation, as obtained by EDX analysis.**

In all samples, the Ti content in the main (matrix) phase ranges from 49 at% to 52 at%, while the sum of the Fe and X contents are between 49 at% and 51 at%. As such, this matrix phase can be attributed to the TiFe phase. The compositional balance between Fe and X indicates that X substitutes Fe in the TiFe structure, in line with previous reports [26].

In Co-based alloy, a secondary phase with a high Ti content (72 at%) was observed, that, from the ternary phase equilibrium diagram, suggests the presence of Ti$_2$(Fe,Co) phase, essentially a Ti$_2$Fe phase where Fe is partially replaced by Co [27].

When Cr was added, a secondary phase with a significant Cr content formed, with 17 at% Cr in the as-cast sample and 27 at% Cr after annealing. The ratio between Ti:(Fe,Cr) matches that reported in the literature for the Ti(Fe,Cr)$_2$ phase and the phases suggested by the ternary phase equilibrium diagram [28,29].

In the case of Al-based as-cast alloy, the matrix exhibited compositional inhomogeneity, evidenced by variations in composition detected through EDX analysis (**Table 1**) and the differing grey nuances observed in the BSE images (**Figure 1**). The observed compositional inhomogeneity within the TiFe matrix in Al-based as-cast alloy is particularly intriguing when considering the pseudo-binary (**Figure S 2**) and the ternary phase diagram [30]. While the nominal composition of Al-based as-cast alloy falls within a region where only a single TiFe phase is expected at equilibrium, it is located near a boundary where two distinct bcc phases coexist. This proximity raises the possibility that local fluctuations in composition, and minor deviations from equilibrium conditions could result in the partial stabilization of two slightly different bcc phases. Given the rapid cooling conditions inherent to arc melting, it is likely that the system did not fully equilibrate, leading to the formation of compositionally distinct regions within the matrix. This is further supported by the fact that, after annealing, only a single TiFe phase is observed, indicating that the initial inhomogeneity was a non-equilibrium effect that disappeared upon achieving thermodynamic equilibrium. In all samples, the EDX analysis revealed the formation of a secondary phase, which shows the presence of oxygen with a Ti to (Fe,X) ratio of 2:1 that, from results reported in the literature, suggest the presence of a $Ti_4(Fe,X)_2O_x$ phase [31–33].

The compositional analysis also indicated that X is present in the oxide phase. In the case of the as-cast Al-substituted alloy, this oxide phase also displayed two slightly different compositions, as shown in **Table 1**. This inhomogeneity was resolved after the annealing process, resulting in a homogenized matrix and only one composition for the oxide phase.

The XRD patterns of the as-cast and annealed alloy powders (**Figure 2**) was acquired to confirm the presence of the phases observed by EDX analysis. Results of the Rietveld refinement of XRD patterns are reported in **Table 2**.

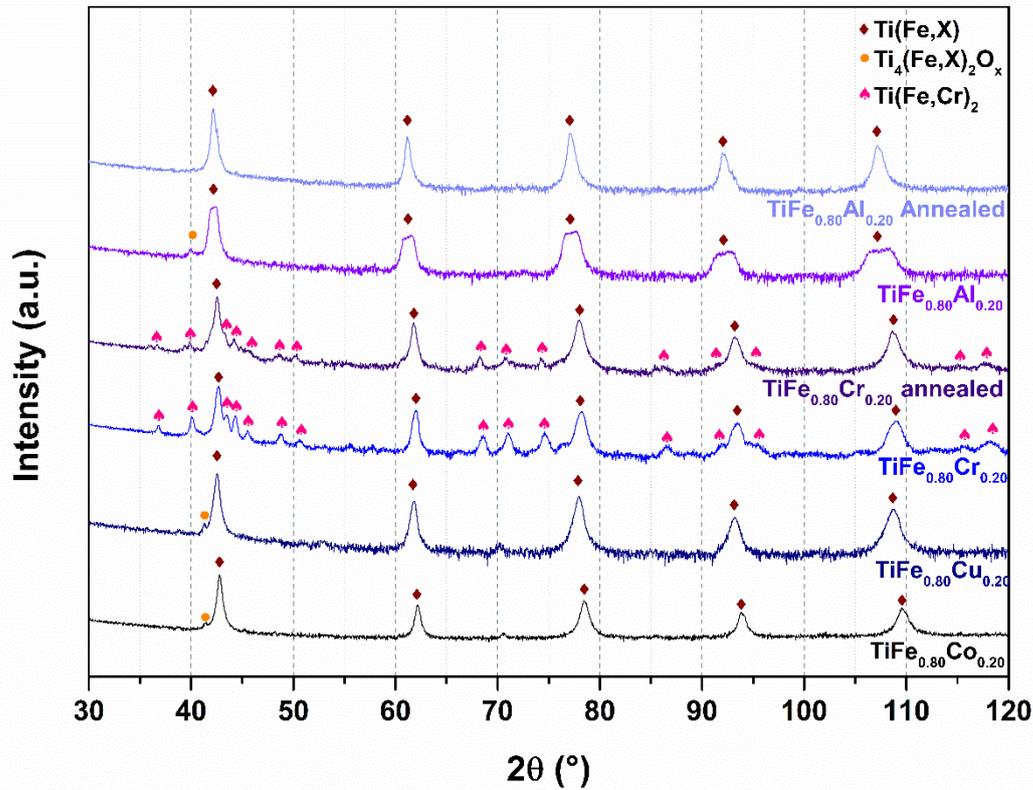

**Figure 2: X-ray diffraction patterns of as cast alloys and the annealed ones. Data is presented with a logarithmic scale on the Y-axis.**

The phases detected by XRD are in good agreement with the analysis performed with the EDX; however, in most cases, their quantities could not be determined by XRD being the collected signal below the detection limit. The exceptions are the phase Ti(Fe,Cr)$_2$ and the oxide phase, which in most cases were present in sufficient amounts to be quantified by XRD (**Table 2**). Therefore the presence of the oxide phase Ti$_4$(Fe,X)$_2$O$_x$ was confirmed. This phase has been documented in the literature with a variable composition, showing no significant changes in the lattice constant [31]. As a result, determining the precise stoichiometry of the oxide phase based on the cell parameter alone is not feasible. Furthermore, the identified Ti(Fe,Cr)$_2$ phase is a C14 Laves phase. From the EDX analysis it is clear that the ratio Ti:(Fe,Cr) matches that reported in the literature for the Ti(Fe,Cr)$_2$ phase [28]. This phase has a Fe:Cr ratio that makes it more likely to be TiFe$_2$ phase rather than TiCr$_2$, due to the higher Fe content compared to Cr. Nevertheless, XRD data suggest that both phases could be present, as $TiFe_2$ and $TiCr_2$ only differ in the shift of their peaks, and the experimental peaks are positioned at a 2θ angle intermediate between those of the two phases.

From the Rietveld analysis of the XRD patterns of the Cr-containing alloys and it is possible to see that, after annealing, the amount of $Ti(Fe,Cr)_2$ phase decreases from 14 wt% to 8 wt% (**Table 2**). This effect is particularly noticeable in the case of the addition of Al, which promotes significant compositional inhomogeneity within the matrix, as evidenced by the double peaks in the pattern. However, this issue was resolved after alloy annealing, which not only promotes matrix homogenization, but also leads to the formation of a single oxide phase with a well-defined composition.

| Sample | TiFe $a$ (Å) | Matrix (wt%) | $Ti_4Fe_2O_x$ (wt%) | $Ti(Fe,Cr)_2$ (wt%) |
|---|---|---|---|---|
| $TiFe_{0.80}Co_{0.20}$ | 2.978± 0.001 | 97.7±0.1 | 2.3±0.3 | - |
| $TiFe_{0.80}Cu_{0.20}$ | 2.997± 0.001 | 99.0±0.1 | 1.0±0.2 | - |
| $TiFe_{0.80}Cr_{0.20}$ | 2.998± 0.001 | 86.0±0.1 | - | 14.0±0.4 |
| $TiFe_{0.80}Cr_{0.20}$ annealed | 2.998± 0.001 | 90.4±0.1 | 1.0±0.2 | 8.6±0.4 |
| $TiFe_{0.80}Al_{0.20}$ | 3.034± 0.001 | 39.1±1.5 | 1.1±0.1 | - |
| | 3.008± 0.001 | 59.8±1.6 | - | - |
| $TiFe_{0.80}Al_{0.20}$ annealed | 3.028± 0.001 | 100.0±0.1 | - | - |

**Table 2: Phase distribution and lattice parameter a of the CsCl-type TiFe phase for all the alloys presented in this work, as determined by Rietveld refinement of XRD data. For the cell parameters of the other phases see Table S 1. *For sample Al-based as-cast alloy, two lines are reported because the matrix is inhomogeneous, resulting in two distinct compositions and corresponding cell parameters, which are presented in separate lines.**

The lattice parameters of the cubic CsCl-type TiFe phase are reported in **Table 2**, and, in the case of Al-based annealed alloy, it is in good agreement with literature values (a=3.015 Å [25]), while, for the other alloys, there is a lack of data in the reference literature. A detailed examination shows a slight variation in the lattice constant of TiFe phase, which is linked to the atomic radius of the substituting element. Specifically, for Co, Cu, and Cr, the lattice parameter increases, ranging from 2.978 Å (Co) to 2.998 Å (Cr). This is consistent with the observation that Co, Cu, Cr, and Ti can substitute Fe in the TiFe structure. In alloys containing Cr and Cu, the lattice parameter increases because these elements, along with Ti, have larger atomic radii than Fe. Consequently, when they substitute Fe, the unit cell expands. The largest cell parameters that of the Al-containing alloy, because of Al greater radius compared to the other substituting elements. After annealing, the cell parameter, 3.028 Å, becomes close to an average of values found for the two slightly different TiFe phases in the as cast alloy, reflecting the compositional homogenization of the matrix [24].

**Hydrogen sorption properties**

**PCI activation**

The first hydrogenation cycle is reported for all the alloys in **Figure 3(a),** with **Figure 3(b)** providing a zoomed view to better illustrate the incubation time. This incubation time refers to the $H_2$ exposure needed before hydrogen sorption begins, and it is determined by observing the onset time of the curve. . It is important to note that the curves in Figure 3 do not present any point precisely at the origin of the graph and this is a consequence of the experimental procedure. Indeed, in the measurement setup, when 64 bar (45 bar in case of Co-based alloy) of hydrogen are introduced into the chamber, there is an initial phase during which the chamber gradually fills and stabilizes at 64 bar before the actual measurement is recorded. During this brief interval, estimated in few seconds, which precedes the start of the measurement, the alloy may already begin to absorb a small amount of hydrogen Consequently, the capacity at time t=0 is not exactly zero and it represents the lag time between the moment when the alloy is first exposed to hydrogen and the measurement start. It corresponds to the time required for the pressure to stabilize before the formal measurement begins. This distinction is critical for accurately interpreting the data and understanding the hydrogen absorption behavior of the alloy from the onset time, and it is particularly evident for Al-based as-cast and Al-based annealed alloys.

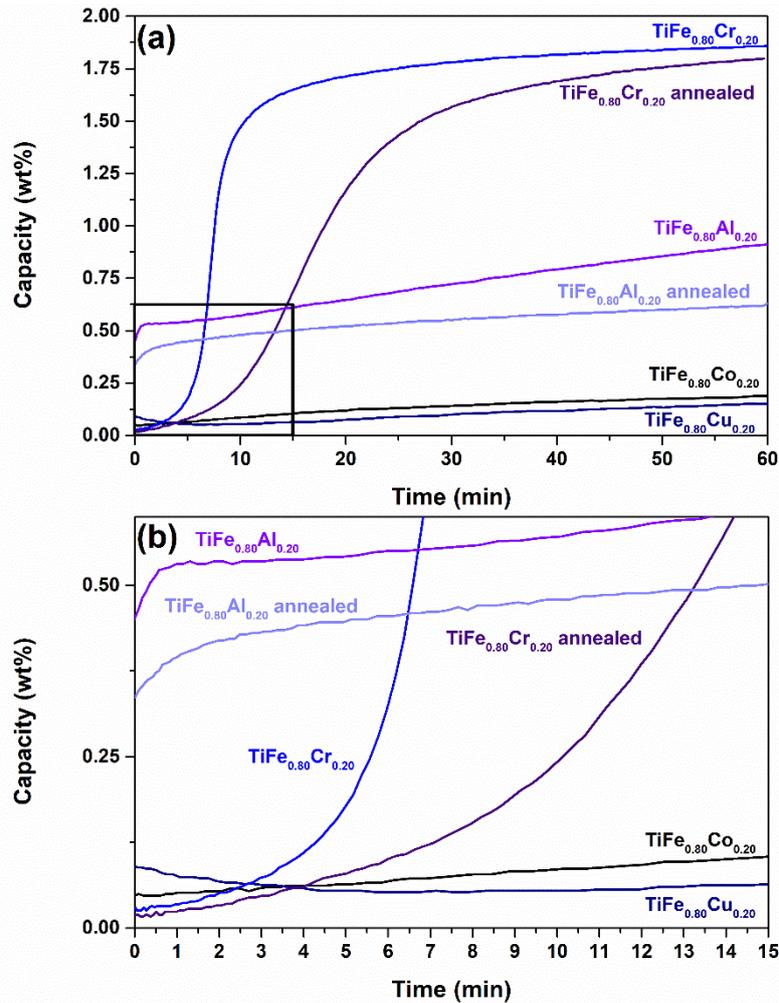

**Figure 3:** First hydrogenation curves of (a) all the alloys at 21 °C and under 64 bar (45 bar for Co-based alloy) hydrogen pressure. (b) Enlargement of the first hydrogen curves, measured at the region marked by the black box in Figure 3(a).

In **Figure 4**, the maximum storage capacity achieved in each cycle is plotted against the $H_2$ exposure time for all alloys, rather than the number of cycles, since different alloys reach their maximum capacity at different times, leading to variations in cycle duration. This figure also illustrates the total activation time, which refers to the overall $H_2$ exposure time needed to reach the first cycle where the sample achieves its maximum storage capacity. For the Cr-substituted alloys, the maximum storage capacity is reached during the first cycle, but the true reversible storage capacity is realized in the second cycle. As will be described in more detail later, this initial drop in capacity is due to the irreversible absorption of H$_2$ by TiCr$_2$ phase. Therefore, for both alloys, activation is considered complete at the end of the second cycle.

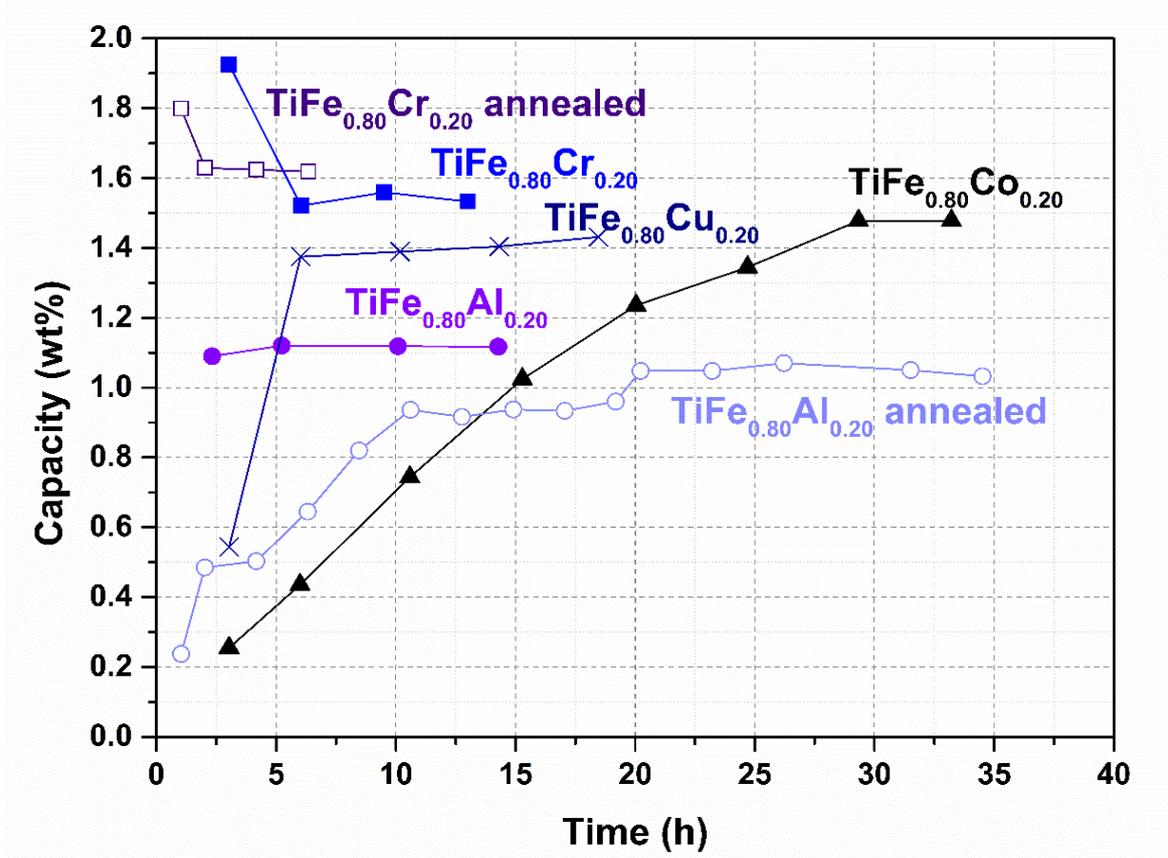

**Figure 4: Maximum storage capacity in each cycle for all alloys as a function of the activation time. Each point represents the last values of each absorption cycle, and so the highest H$_2$ capacity reached in each absorption cycle.**

If we consider the incubation time (**Figure 3** and

**Table** 3), both Al-based samples showed a low incubation time (instant absorption), followed by Cr-based alloys (1-2 min), Cu-based alloy (5 min) and Co-based alloy (12 min). If the total time for activation is considered (**Figure 4** and

**Table** 3), Al-based as-cast alloy and Cr-based samples fully activate in less time (5 and 2-6 h, respectively) than the other alloys, followed by Cu (10 h), Al (20 h), and finally, Co (29 h) based alloys.

|  | P (bar) | %H (wt%) | T (°C) | Inc t (min) | t(h) |
| --- | --- | --- | --- | --- | --- |
| **TiFe$_{0.80}$Co$_{0.20}$** | 45 | 1.5 | 22 | 12 | 29 |
| **TiFe$_{0.80}$Cu$_{0.20}$** | 65 | 1.4 | 20 | 5 | 10 |
| **TiFe$_{0.80}$Cr$_{0.20}$** | 62 | 1.5 | 23 | 1 | 6 |
| **TiFe$_{0.80}$Cr$_{0.20}$ annealed** | 61 | 1.7 | 23 | 2 | 2 |
| **TiFe$_{0.80}$Al$_{0.20}$** | 65 | 1.1 | 21 | 0 | 5 |
| **TiFe$_{0.80}$Al$_{0.20}$ annealed** | 63 | 1.1 | 24 | 0 | 20 |

**Table 3: Data for the activation at room temperature and with H$_2$ (Sievert apparatus). Inc t is the incubation time, which is the duration needed to initiate hydrogen absorption. t represents the minimum exposure time to H$_2$ required to achieve full activation.**

Among all alloys, Cr-based alloys are the easiest to fully activate, likely due to the presence of the TiCr$_2$ phase, which is known to readily absorb hydrogen [34], thereby facilitating activation. That is also why the as-cast Cr-based alloy, which has more Ti(Fe,Cr)$_2$ phase compared to the annealed alloy (14 wt% and 8 wt% respectively, **Table 2**) has a slightly lower incubation time (1 min and 2 min, respectively,

**Table** 3) to start hydrogen absorption. Thus, in multiphase alloys, the Ti(Fe,Cr)$_2$ phase preferentially hydrogenates over TiFe. Hydrogenation of the secondary phase causes volume expansion, generating numerous cracks and exposing fresh surfaces in the matrix phase, that serve as pathways for hydrogen migration [21]. Therefore, secondary phases present at grain boundaries can be easily reached by hydrogen atoms by diffusion, enabling their hydrogenation, with a consequent change in volume and, eventually, the brake of the alloy powder with a detachment of the grains and an exposure of fresh surface ready to be hydrogenated [35–37]. It is therefore of importance to take into account the presence of secondary phases at the grain boundaries, that can easily absorb H$_2$ in the activation process, due to their evident aid in the initiation of the activation process. In the case of the studies alloys, the Ti(Fe,Cr)$_2$ phase can be considered as a facilitator for fast activation [34].

A similar mechanism applies to other alloys, though they lack the presence of secondary phases. However, the as-cast alloy Al-based alloy exhibits significant compositional heterogeneity, creating defects which can act as fast diffusion channels for hydrogen. In the absence of a high hydrogen-affinity phases, such as $TiCr_2$, the incubation time is longer than for Cr-containing alloys (1-2 min for as-cast and annealed alloys, respectively, **Figure 3** and **Table** 3). In the case of both Al-based samples, there is no phase like $TiCr_2$ that promotes hydrogen absorption. However, they exhibit no incubation time (**Table** 3), as they begin absorbing hydrogen immediately upon contact with it. In fact, for the as-cast alloy, as mentioned earlier, compositional inhomogeneity plays a crucial role. In the case of the annealed alloy, the same behavior occurs, despite its compositional homogeneity. It is likely that the initiation of hydrogen absorption is facilitated by the addition of Al due to its electronic effects. As extensively discussed in the work of Nambu et al. [38], the amount of hydrogen stored in the TiFe-X alloy is influenced by the strength of the Ti-X bond. A weaker bond results in a lower quantity of hydrogen being stored, while a stronger bond correlates with a higher hydrogen storage capacity. Additionally, it is noted that, in the period of the table of elements, the bond strength decreases with the atomic number of element X [38]. Based on these considerations, along with the atomic numbers of X and the findings from this study, as well as the bond energies of the Ti-X [39], it can be deduced that the Ti-Al bond is stronger than the others. This stronger bond, in turn, favors the absorption of hydrogen, as demonstrated by Nambu et al [38].

In the other alloys that lack a high-affinity secondary phases and have matrix's high compositional homogeneity, incubation times are significantly longer (5-15 min for Cu-based alloy and Co-based alloy alloys respectively, **Table** 3).

A key factor to consider in TiFe-based alloys activation, particularly first hydrogenation and incubation time, is the presence of a passivating surface oxide layer. In fact, TiFe-based alloys inherently contain ternary oxides, due to Ti high affinity for oxygen and the immiscibility of oxygen in the TiFe-based alloy matrix [31]. As commonly stated in the literature, in TiFe based alloys, a $Ti_3Fe_3O$ oxide layer forms on the surface, creating a passivating oxide layer that is unable to absorb hydrogen, thereby hindering the first hydrogenation. As a result, high temperatures and hydrogen pressures are generally required to promote hydrogen diffusion through this passivating layer [20,40–42]. Once this oxide layer is penetrated, the presence of grain boundaries between oxides and the matrix serves as fast diffusion pathways for hydrogen, facilitating the hydrogenation of the matrix, which expands, while the surface oxide layer remains unchanged, due to its inability to absorb hydrogen, as previously mentioned. This volumetric expansion difference, along with hydrogen absorption in the $Ti_4Fe_2O_x$ phase, lead to crack formation, creating new surfaces that further promote hydrogenation [43–46].

Thus, the incubation time can essentially be interpreted as the time required for hydrogen to penetrate this passivating oxide layer. Once hydrogen has diffused through this layer, the variation in incubation time among different alloys is primarily determined by the presence of additional grain boundaries which arise from the presence of secondary phases (e.g., Cr-substituted alloys) or regions with different compositions (e.g., Al-substituted alloys), which further enhance hydrogen diffusion. This effect is even more pronounced if phases capable of absorbing hydrogen are present, as in the case of Cr-substituted alloys [47]. This distinction may contribute to the differences in incubation time observed among Al-, Cr-, Co-, and Cu-substituted alloys.. Considering the process to reach the full activation, diffusion processes within the bulk play an important role, as demonstrated by numerous studies showing their positive influence on activation kinetics [48–50].

Secondary phases within the matrix create multiple hydrogen diffusion pathways, since grain boundaries are known to enhance hydrogen diffusivity within alloys. In Cr-substituted alloys the abundance of grain boundaries and high hydrogen-affinity secondary phases creates numerous fast hydrogen diffusion channels in the bulk, favoring both hydrogen absorption kinetics and alloy activation [51]. A similar scenario applies to the Al-based as cast alloy. In contrast to the annealed alloy, it has a high degree of compositional heterogeneity (**Figure 1, Table 1**Table 1

| Sample | Average Composition | | | Phase | Phase Composition | | |
|---|---|---|---|---|---|---|---|
| | Ti (at%) | Fe (at%) | X (at%) | | Ti (at%) | Fe (at%) | X (at%) |
| TiFe0.80Co0.20 | 51.3±0.2 | 39.0±0.1 | 9.7±0.2 | TiFe | 51.3±0.3 | 38.5±0.9 | 10.2±1.1 |
| | | | | Ti4Fe2Ox | 60.7±1.7 | 33.7±1.2 | 5.7±0.5 |
| | | | | Ti2(Fe,Cr) | 72.3±5.3 | 23.5±4.6 | 4.3±0.9 |
| TiFe0.80Cu0.20 | 51.1±0.1 | 39.5±0.6 | 9.4±0.5 | TiFe | 50.8±0.2 | 40.6±0.3 | 8.7±0.4 |
| | | | | Ti4Fe2Ox | 63.0±1.8 | 29.4±1.4 | 7.6±0.8 |
| TiFe0.80Cr0.20 | 51.2±0.2 | 38.4±0.5 | 10.4±0.7 | TiFe | 52.4±0.4 | 40.7±0.6 | 6.9±0.9 |
| | | | | Ti4Fe2Ox | 69.0±2.0 | 21.4±1.9 | 9.6±0.3 |
| | | | | Ti(Fe,Cr)₂ | 44.2±1.1 | 39.2±1.8 | 16.6±0.9 |
| TiFe0.80Cr0.20 Annealed | 51.0±0.2 | 39.1±0.2 | 9.9±0.1 | TiFe | 51.6±0.1 | 40.2±0.2 | 8.2±0.3 |
| | | | | Ti4Fe2Ox | 61.3±1.8 | 29.7±2.4 | 9.0±0.6 |
| | | | | Ti(Fe,Cr)₂ | 41.6±0.6 | 31.7±0.8 | 26.7±1.4 |
| TiFe0.80Al0.20 | 50.6±0.1 | 37.8±0.1 | 11.6±0.1 | TiFe (1) | 48.7±1.0 | 41.1±2.3 | 10.2±1.3 |
| | | | | TiFe (2) | 50.3±1.1 | 37.2±3.0 | 12.4±2.2 |
| | | | | Ti4Fe2Ox (1) | 61.8±3.1 | 25.8±2.6 | 12.4±0.5 |
| | | | | Ti4Fe2Ox (2) | 69.3±1.5 | 16.7±1.3 | 14.0±0.2 |
| TiFe0.80Al0.20 Annealed | 50.5±0.2 | 37.8±0.1 | 11.8±0.1 | TiFe | 50.5±0.1 | 39.5±0.2 | 10.1±0.2 |
| | | | | Ti4Fe2Ox | 65.6±0.3 | 22.8±0.3 | 11.7±0.3 |

) that promotes the formation of defects, thereby enhancing the diffusion of hydrogen and accelerating complete activation. Conversely, the remaining alloys (i.e. Al annealed, Co and Cu-based alloys) show compositional homogeneity, with few or no secondary phases, reinforcing the interpretation that the limited number of fast hydrogen diffusion channels leads to a slower overall activation process.

The individual elements substituting Fe in the TiFe alloy should also be considered. Various studies indicate that each added element (in this case, Co, Cu, Cr, or Al) can significantly impact hydrogen diffusion kinetics, the diffusion coefficient, and the activation energy for hydrogenation, due to their unique electronic effects and affinities for hydrogen [48,49,51]

A recent study by Bakulin et al. [51] showed that elements such as Co, when doped into the TiFe alloy, lower the hydrogen diffusion coefficient and increase the activation energy for hydrogen diffusion. By contrast, Cu increases the hydrogen diffusion coefficient but raises the activation energy. On the other hand, elements such as Al and Cr not only increase the hydrogen diffusion coefficient but also lower the activation energy for diffusion. This is reflected in the trends seen in **Figure 4** for the alloys analyzed here, where Cr-based alloys and Al-based as-cast alloy fully activate more easily, followed by Cu and Co-based alloys.

Another factor to consider is the lattice parameter of the TiFe-based phase, calculated for each alloy, and discussed in the XRD characterization section (**Figure 2, Table 2**). Although the lattice parameter remains relatively consistent across the alloys, it is slightly larger in Al-containing alloys, which may improve activation. According to Alefeld and Völkl's theory [48], an increase in the matrix's specific volume due to doping can decrease the activation energy for hydrogen absorption and accelerate hydrogenation kinetics. This reduction in activation energy is associated with increased interstitial volume, that can also influence the interstitial diffusion coefficient. As for the Al-substituted alloy, after annealing, its composition becomes homogeneous with a cell parameter that is in between the two original values. That means that, possibly, the Al-based as cast alloy starts its hydrogenation by taking advantage of the phase with higher cell parameter, while, after the annealing, the smaller cell dimension reduces the hydrogen diffusion coefficient hindering the absorption process.

It should also be noted that Cr-substituted alloys show a significant drop in capacity after the first absorption cycle. This is likely due to hydrogen absorption by the $TiCr_2$ phase, which forms stable hydrides, so they do not release hydrogen during the first desorption cycle, resulting in slightly lower capacity in subsequent cycles. The capacity drop is more noticeable in the case of Cr-based as-cast alloy, which contains a higher quantity of the $Ti(Fe,Cr)_2$ phase (14 wt%) compared to the annealed alloy (8 wt%). As previously mentioned, this phase does not release hydrogen, therefore due to the higher proportion of $Ti(Fe,Cr)_2$ in the as cast alloy, the capacity drop after the first absorption cycle is higher (from 1.9 to 1.5 wt%) than in the annealed alloy (from 1.8 to 1.6 wt%). Consequently, the final maximum capacity of the as cast alloy (1.5 wt%) turns out lower than that of the annealed one (1.6 wt%). Furthermore, the capacity of Al-based alloys is significantly lower (1.0-1.1 wt% for the annealed and as cast alloys, respectively) than other alloys. This is because the addition of Al impedes the formation of the γ dihydride phase, as documented in the literature [25]. Since only the β monohydride phase forms, the maximum storage capacity is inevitably lower than that of other alloys. In summary, the activation process is influenced by both the interaction between hydrogen and the surface, which affects the incubation time, and the diffusion of hydrogen inside the bulk, which is more correlated to the total time of the activation. Therefore, to have a clear picture of the hydrogenation mechanism, it is essential to consider all parameters that influence the process. This encompasses secondary phases with high hydrogen affinity and the presence of hydrogen diffusion pathways (e.g., grain boundaries, phase boundaries), as well as parameters like composition, lattice parameters, and the hydrogen diffusion coefficient.

**HPDSC activation**

Results of HPDSC activation are reported in **Figure 5** and data are summarized in **Table 4**. It is important to note that acquisition times considered in **Table 4** are calculated by taking into account the entire activation process, starting from t=0, which corresponds to the beginning of the isothermal step. Here, Inc t refers to the incubation time, defined as the onset time of the first exothermic peak (absorption peak) observed during the isothermal step and first heating cycle. The time t represents the duration required to achieve full activation. It is determined based on the offset time of the first reproducible exothermic peak. To be reproducible this peak has to remain reproducible in all subsequent cycles, serving as a reliable indicator of activation completion.

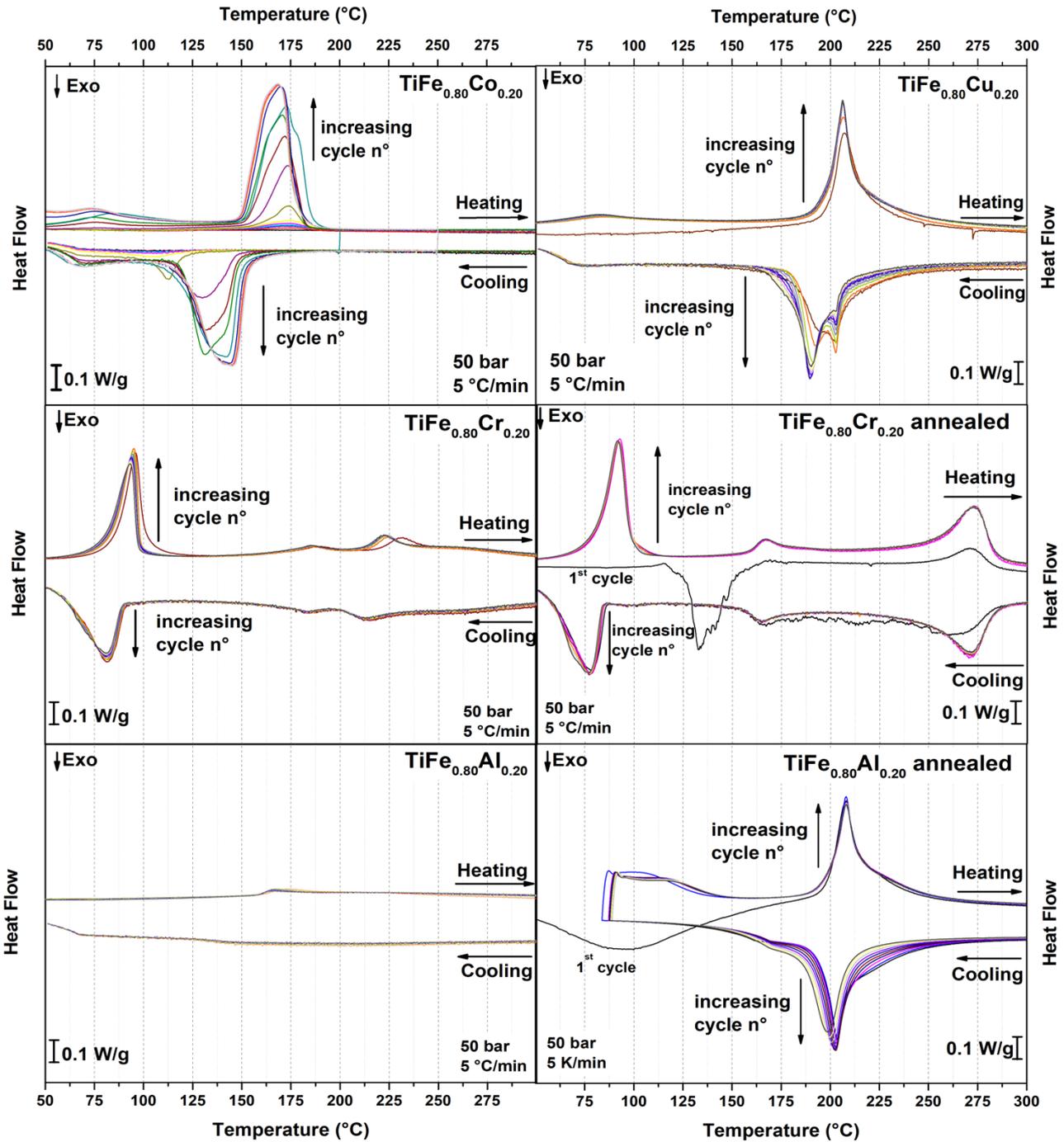

**Figure 5:** DSC curves of all alloys at 50 bar of H$_2$ pressure. The Heat Flow is reported as a function of Temperature.

|  | Inc t (h) | t (h) | Pmin (bar) | Pmax (bar) | T$_{p1}$ (°C) | T$_{p2}$ (°C) | T$_{p3}$ (°C) |
|---|---|---|---|---|---|---|---|
| **TiFe$_{0.80}$Co$_{0.20}$** | 5 | 15 | 52 | 65 | 76 | 168 |  |
| **TiFe$_{0.80}$Cu$_{0.20}$** | 1 | 6 | 52 | 66 | 82 | 207 |  |
| **TiFe$_{0.80}$Cr$_{0.20}$** | 0.1 | 4 | 52 | 66 | 95 | 185 | 223 |
| **TiFe$_{0.80}$Cr$_{0.20}$ annealed** | 3 | 4 | 50 | 65 | 91 | 168 | 272 |

| | | | | | | |
|---|---|---|---|---|---|---|
| TiFe$_{0.80}$Al$_{0.20}$ | 3 | 4 | 52 | 67 | 167 | |
| TiFe$_{0.80}$Al$_{0.20}$ annealed | 3 | 4 | 51 | 69 | 118 | 208 |

**Table 4. Activation data for HPDSC experiments for all samples. Inc t is the incubation time, which is the duration needed to initiate hydrogen absorption, and it is calculated considering both the isotherm and first heating cycle. t represents the minimum exposure time to H$_2$ required to achieve full activation, and it is calculated from the first reproducible heating cycle. T$_{p1}$, T$_{p2}$, and T$_{p3}$ represent the temperatures at which the maximum of each desorption peak occurs, during the last heating cycle.**

In the two annealed alloys, one curve (1$^{st}$ cycle) is visibly different with a first exothermic peak (absorption peak).

Only Cu-based alloy and Cr-based as-cast alloy exhibit a clear exothermic peak related to hydrogen absorption during the isothermal step at 40 °C (**Figure S 1**). For Cu-based alloy, the peak is less pronounced compared to Cr-based as-cast alloy, but the isothermal curve is not completely flat; towards the end of the 3-h period, a slight slope is observed due to hydrogen absorption. As a result, these alloys are the first to begin activation, with incubation times of 1 h and 7 min, respectively. They are followed by the Cr-based annealed, Al-based as-cast and Al-based annealed alloys, which activate during the first heating cycle, with an incubation time of 3 h, after the temperature increase. These alloys display an exothermic absorption peak during the first heating cycle, followed by an endothermic desorption peak. In contrast, Co-based alloy remains the most difficult to be activated, with an incubation time of 5 h. It does not exhibit either exothermic absorption peaks or endothermic desorption peaks during the isothermal step or the first heating cycle.

**Comparison of activation in PCI and HPDSC**

It is noteworthy that, when a lower hydrogen pressure is used (50 bar compared to 64 bar in the Sievert apparatus), all alloys are more difficult to be activated, showing longer incubation times. In PCI, the incubation time is only a few minutes, whereas in the HPDSC, it takes several hours (with the exception of the Cr-based as-cast alloy, although its incubation time is still longer than that at 64 bar in the Sievert apparatus). Therefore, with the decrease in hydrogen pressure from 64 bar to 50 bar, hydrogen absorption starts with a longer incubation period, and for some alloys, an increase in temperature is also necessary.

Indeed, both pressure and temperature conditions influence the driving force $\Delta G$ for the hydride phase formation [52], as well as the nucleation frequency [53,54]. Considering these parameters, lowering the hydrogen pressure, as in the HPDSC measurements, disfavors the absorption, by reducing $\Delta G$. At the same time, the decrease in pressure decreases the nucleation frequency. This explains why, at lower pressure as in our case, the incubation times are prolonged.

Regarding the increase in temperature required for some alloys, an increase in temperature reduces the thermodynamic driving force for hydride formation, by making $\Delta G$ less negative. However, it simultaneously increases the $H_2$ diffusion coefficient, thereby promoting the nucleation and growth kinetics of the hydride phase. This further explains why, in terms of the time required to reach the maximum storage capacity — i.e., the full activation of the alloys — times measured with the HPDSC are shorter compared to those measured in the PCI (except for the Cr-based annealed alloy). Thus, an increase in temperature enhances the kinetics of the absorption process (i.e. the diffusion of hydrogen into the alloy) and leads to shorter activation times.

## Conclusions

The present study has provided detailed insights into the activation process of TiFe-X alloys, with a focus on the role of the third element (X) in influencing hydrogen absorption behavior, activation kinetics, as well as structural and microstructural properties. The chemical and structural analyses revealed, in some cases, the formation of secondary phases, such as $Ti_2(Fe,Co)$, $Ti(Fe,Cr)_2$, and $Ti_4(Fe,X)_2O_x$, which are critical in determining the activation behavior of these alloys. These phases, particularly those rich in Cr, facilitate hydrogen absorption, by providing additional pathways for hydrogen diffusion, thereby reducing the incubation time required for activation.

The hydrogen sorption properties, as evaluated by PCI and HPDSC, demonstrated that alloys containing Cr and Al exhibited shorter activation times, compared to other compositions. This has been attributed to the high hydrogen affinity of Cr-containing phases, such as $TiCr_2$ [34], and the compositional heterogeneity present in the Al-substituted alloy, which promotes the formation of numerous grain boundaries, that serve as fast diffusion channels for hydrogen. Furthermore, the adding of these two elements also improves the diffusion of hydrogen inside the lattice [51].

The study also highlighted the influence of the lattice parameter of the FeTi-based phase on activation. Elements such as Cr and Al, which increase the lattice parameter, contribute to more favorable conditions for hydrogen absorption, by enhancing the interstitial volume, which in turn lowers the activation energy for hydrogenation and accelerates diffusion processes. This finding aligns with previous studies, suggesting that an increase in lattice parameter can positively affect activation kinetics by providing more space for hydrogen atoms to diffuse [48].

In addition, the comparison of activation behavior in different experimental setups revealed that both pressure and temperature play crucial roles in influencing the activation process, as already discussed in Yartys et al. [52]. Lower hydrogen pressure and increased temperature were found to extend the incubation time, but improve the overall activation kinetics, particularly for alloys without high-affinity secondary phases.

In conclusion, the activation process of TiFe-X alloys is strongly dependent on the presence of specific secondary phases, the lattice parameter, and the diffusion pathways within the matrix. The incorporation of elements like Cr and Al reduces the activation time. Future work will focus on further optimizing these alloys, considering both their structural properties and hydrogen absorption behavior, to achieve more efficient and reproducible hydrogen storage systems.

# CRediT authorship contribution statement

**Francesca Garelli**: Conceptualization, Data curation, Formal analysis, Investigation, Methodology, Validation, Visualization, Writing - original draft, Writing - review & editing. **Erika Michela Dematteis**: Conceptualization, Data curation, Formal analysis, Investigation, Methodology, Validation, Visualization, Supervision, Funding acquisition, Project administration, Writing - review & editing. **Vitalie Stavila, Giuseppe Di Florio, Claudio Carbone, Alessandro Agostini, Mauro Palumbo**: Conceptualization, Formal analysis, Validation, Writing - review & editing, Project administration. **Marcello Baricco, Paola Rizzi**: Conceptualization, Funding acquisition, Resources, Supervision, Validation, Visualization, Project administration, Writing - review & editing.

# Declaration of Competing Interest

The authors declare that they have no known competing financial interests or personal relationships that could have appeared to influence the work reported in this paper.

# Supplementary data

Dataset related to the publication and raw data are available online at https://doi.org/10.5281/zenodo.15051094 and https://doi.org/10.5281/zenodo.15051336 [55,56].

# Acknowledgement

The authors want to acknowledge the **EX-MACHINA** project leading to this publication, which it has received funding under the MUR program "PNNR M4C2 Initiative 1.2: Young Researcher - Seal of Excellence" (CUP: D18H22002040007), the support from Project CH4.0 under MUR program "**Dipartimenti di Eccellenza 2023-2027**" (CUP: D13C22003520001) and the support received under the National Recovery and Resilience Plan (NRRP), Mission 4 Component 2 Investment 1.4 -Call for tender No. 3138 of December 16, 2021 of the Italian Ministry of University and Research, financed by the European Union NextGenerationEU [Award Number: National Sustainable Mobility Center CN00000023, named **MOST**, Concession Decree No. 1033 of June 17, 2022, adopted by the Italian Ministry of University and Research, Spoke 14 "Hydrogen and New Fuels"].

The authors also acknowledge financial support under the National Recovery and Resilience Plan (NRRP), Mission 4, Component 2, Investment 1.1, Call for tender No. 1409 published on 14.9.2022 by the Italian Ministry of University and Research (MUR), funded by the European Union – NextGenerationEU– Project Title Novel hEat REcovery solutions on board of FC equipped vessels for metal HYDridES storage optimal management (**NEREHYDES**) – CUP B53D23006590006 - Grant Assignment Decree No. 961 adopted on 30/06/2023 by the Italian Ministry of Ministry of University and Research (MUR).

This research was funded by the European Union – NextGeneration EU from the Italian Ministry of Environment and Energy Security POR $H_2$ AdP MASE/ENEA with involvement of CNR and RSE, PNRR - Mission 2, Component 2, Investment 3.5 "**Ricerca e sviluppo sull'idrogeno**", CUP: I83C22001170006.

This publication is part of the **REMEDHYS** project that has received fundings from the European Union's Horizon Europe research and innovation program and supported by the Clean Hydrogen Partnership and its members under grand agreement No 10119203.

# ESI

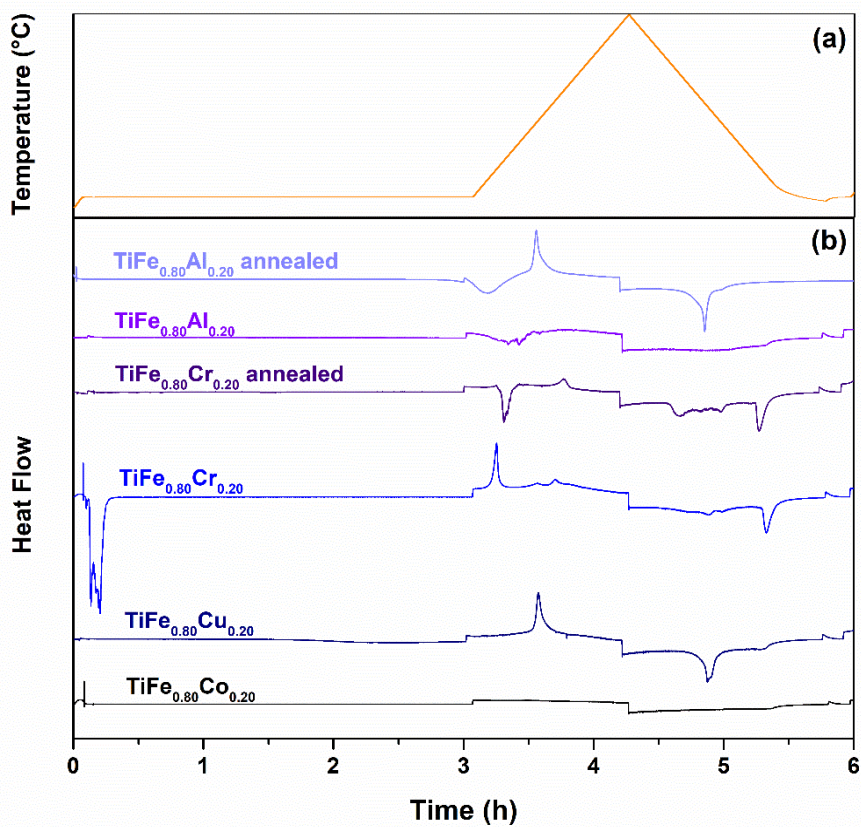

**Figure S 1: First 6 h of the DSC Temperature (a) and of the DSC Heat Flow (b) as a function of the time of all alloys. The heating rate is 5°C/min and the pressure of hydrogen used is 50 bar.**

| Sample | $Ti_4Fe_2O_x$ $a$ (Å) | $TiCr_2$ $a$ (Å) | $TiCr_2$ $c$ (Å) |
|---|---|---|---|
| **$TiFe_{0.80}Co_{0.20}$** | 12.369±0.001 | - | - |
| **$TiFe_{0.80}Cu_{0.20}$** | 11.337±0.001 | - | - |
| **$TiFe_{0.80}Cr_{0.20}$** | - | 4.883±0.001 | 15.946±0.001 |
| **$TiFe_{0.80}Cr_{0.20}$ annealed** | 11.290±0.001 | 4.883±0.001 | 16.037±0.001 |
| **$TiFe_{0.80}Al_{0.20}$** | 11.665±0.001 | - | - |
| **$TiFe_{0.80}Al_{0.20}$ annealed** | - | - | - |

**Table S 1: Lattice parameters of the secondary phases for all alloys as determined by Rietveld refinement of XRD data.**

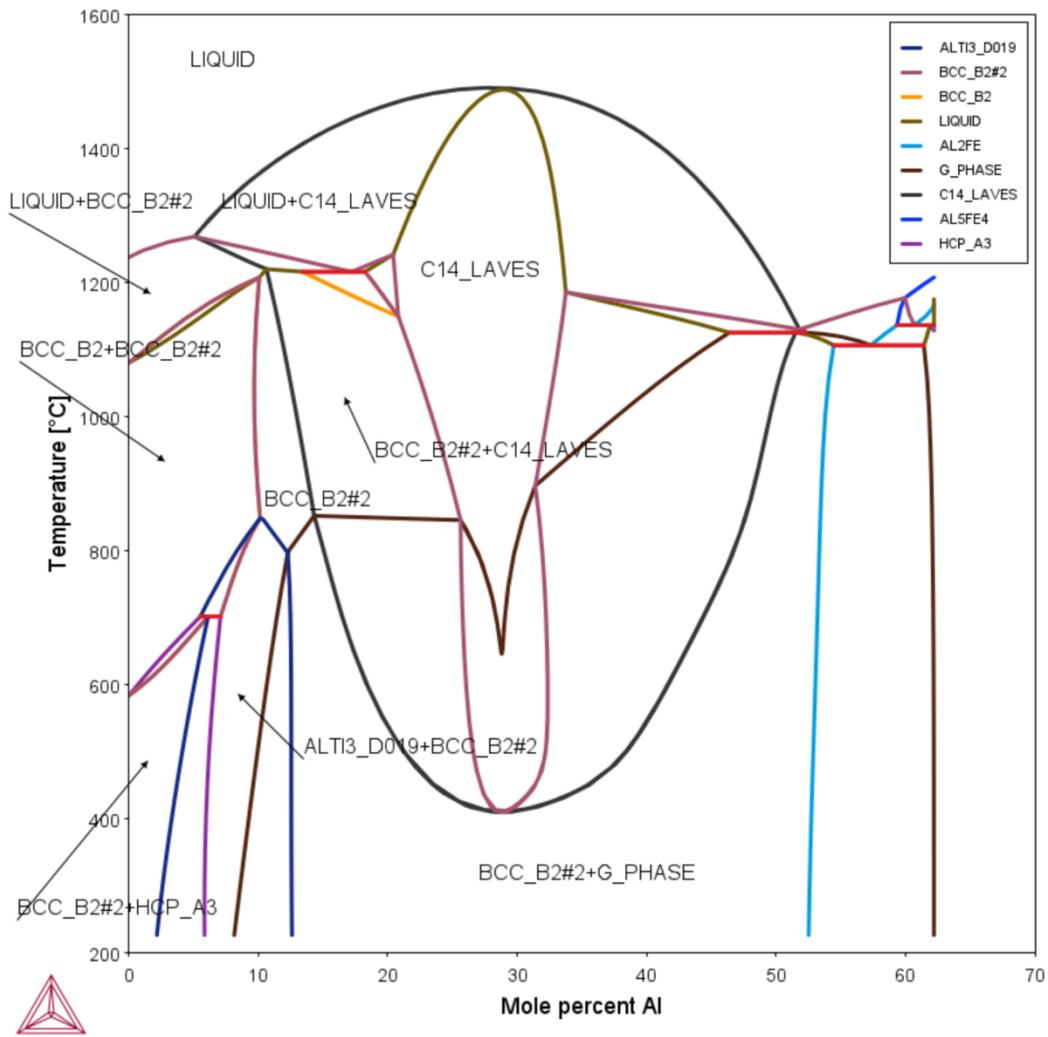

**Figure S 2: Pseudo-binary phase diagram for Ti-Fe-Al system calculated with Thermo-calc software using High Entropy Alloys database.**